\documentclass{ws-p10x7}

\def\MET{\mbox{${\hbox{$E$\kern-0.6em\lower-.1ex\hbox{/}}}_T$}} 
\def\D0{D\O}                            

\begin{document}

\title{Mini-Review on Extra Dimensions}

\author{Greg Landsberg}

\address{Brown University, Department of Physics, 182 Hope St., Providence, RI 02912,
USA\\E-mail: landsberg@hep.brown.edu}

\twocolumn[\maketitle\abstract{
One of the most stimulating recent ideas in particle physics involves a possibility that our universe has additional compactified spatial dimensions, perhaps as large as 1~mm. In this mini-review, we discuss the results of recent experimental searches for such large extra dimensions.}]

The possibility that the universe has more than three spatial dimensions has been a long-discussed issue.\cite{Riemann} Developments in string theory suggest that there could be up to $n=7$ additional dimensions, compactified at very small distances on the order of $10^{-32}$~m. In a new model,\cite{ADD} inspired by string theory, several of the compactified extra dimensions are suggested to be as large as 1~mm. These large extra dimensions (LED) are introduced to solve the hierarchy problem of the standard model (SM) by lowering the Planck scale, $M_{\rm Pl}$, to the TeV energy range. We refer to this effective Planck scale as $M_S$.

Since Newton's law of gravity is modified in the presence of compactified extra dimensions for interaction distances below the size of the LED, current gravitational observations rule out possibility of only a single LED. Recent preliminary results from gravity experiments at submillimeter distances,\cite{tabletop} as well as cosmological constraints,\cite{cosmology} indicate that the case of $n = 2$ is likely ruled out as well. However, for $n \ge 3$, the size of the LED becomes microscopic and therefore eludes the reach of direct gravitational measurements or cosmological constraints. However, high energy colliders, capable of probing very short distances, can provide crucial tests of the LED hypothesis, in which effects of gravity are enhanced at high energies due to accessibility of numerous excited states of graviton, corresponding to multiple winding modes of the graviton field around the compactified dimensions.

LED phenomenology at colliders has already been studied in detail.\cite{GRW,HLZ,Peskin,Hewett} One of the primary observable effects is an apparent non-conservation of momentum caused by the direct emission of  gravitons, which leave the three flat spatial dimensions. A typical signature is the production of a single jet or a vector boson at large transverse momentum. The other observable effect is the anomalous production of fermion-antifermion or diboson pairs with large invariant mass stemming from the coupling to virtual gravitons. Direct graviton emission is expected to be suppressed by a factor $(1/M_S)^{n+2}$, while virtual graviton effects depend only weakly on the number of extra dimensions.\cite{GRW,HLZ,Hewett} Virtual graviton production therefore offers a potentially more sensitive way to search for manifestations of LED.\footnote{Strictly speaking, virtual graviton effects are sensitive to the ultraviolet cutoff required to keep the divergent sum over the graviton modes finite.\protect\cite{GRW,HLZ,Hewett} This cutoff is expected to be of the order of the effective Planck scale. Dependence on the value of the cutoff is discussed, e.g., in Refs.\protect\cite{Shrock,KCGL}}

The effects of direct graviton emission, including production of single photons or $Z$'s, were sought at LEP.\cite{common,direct,L3ZZold} The following signatures were used: $\gamma\MET$ or $Z(jj)\MET$, where \MET\ is the missing transverse energy in the detector, and $j$ stands for jet. The former topology is typical of searches for GMSB supersymmetry, and the latter of searches for ``invisible'' Higgs. The negative results of these searches can be expressed in terms of limits on the effective Planck scale, as summarized in Table~\ref{direct}. Both the CDF and D\O\ Collaborations at the Fermilab Tevatron are also looking for direct graviton emission in the ``monojet'' ($j\MET$) channel, which is quite challenging due to large instrumental background from jet mismeasurement and cosmic rays. Although no results have been reported as yet, the sensitivity of these searches is expected to be similar to those at LEP.\cite{Maria}.

While the formalism for calculating direct graviton emission is well established, different formalisms have been used to describe virtual graviton effects.\cite{GRW,HLZ,Hewett} Since difermion or diboson production via virtual graviton exchange can interfere with the SM production of the same final state particles, the cross section in the presence of LED is given by:\cite{GRW,HLZ,Hewett} $\sigma = \sigma_{\rm SM} + \sigma_{\rm int}\eta_G + \sigma_G\;\eta_G^2,$ where $\sigma_{\rm SM}$, $\sigma_{\rm int}$, and $\sigma_G$ denote the SM, interference, and graviton terms, and the effects of ED are parametrized via a single variable $\eta_G = {\cal F}/M_S^4$, where ${\cal F}$ is a dimensionless parameter of order unity, reflecting the dependence of virtual graviton coupling on the number of extra dimensions. Several definitions exist for ${\cal F}$:
\begin{eqnarray}
	{\cal F} & = & 1, \mbox{~(GRW\cite{GRW});} \nonumber\\
	{\cal F} & = & \left\{ \begin{array}{ll} 
         \log\left( \frac{M_S^2}{M} \right), & n = 2 \\
	   \frac{2}{n-2}, & n > 2
	   \end{array} \right. , \mbox{~(HLZ\cite{HLZ});} \nonumber\\
	{\cal F} & = & \frac{2\lambda}{\pi} = \pm\frac{2}{\pi}, \mbox{~(Hewett\cite{Hewett}).} \nonumber
\end{eqnarray}
Here, $\lambda$ is a dimensionless parameter of order unity, conventionally set to be either $+1$ or $-1$ in cross section calculations within Hewett's formalism. Only the HLZ formalism has ${\cal F}$ depending explicitly on $n$.

Because different experiments have set limits on virtual graviton exchange using different formalisms, it is worthwhile to specify relationship between the three definitions of effective Planck scale, referred to as
$\Lambda_T$, after the original\cite{GRW} notation, $M_S({\rm Hewett})$, and $M_S({\rm HLZ})$:
\begin{eqnarray}
	M_S({\rm Hewett})\left|_{\lambda=+1}\right. & = &
  \sqrt[4]{\frac{2}{\pi}} M_S({\rm HLZ})\left|_{n=4}\right.\nonumber\\
	\Lambda_T & = & M_S({\rm HLZ})\left|_{n=4}\right. .\nonumber\label{GRW-HLZ}
\end{eqnarray}
Unless noted otherwise, we will express limits on the effective Planck scale in terms of $M_S({\rm Hewett})$, and they all will be given at 95\% CL.

Among the many difermion and diboson final states tested for presence of virtual graviton effects at LEP,\cite{common,L3ZZold,virtual,L3ZZ,OPALZZ} the most sensitive channels involve the dielectron (both Drell-Yan and Bhabha scattering) and diphoton ppocesses.\footnote{
Recent preliminary results from L3 at $\sqrt{s} > 200$~GeV indicate that the best sensitivity is found in the $ZZ$ channel,\cite{L3ZZ} but details of the experimental analysis are not yet available. These results differ from those of an earlier L3 publication,\cite{L3ZZold} where the sensitivity in the $ZZ$ channel at $\sqrt{s} = 189$~GeV was significantly lower than that in the $\gamma\gamma$ channel, as well as from recent OPAL results in the $ZZ$ channel at the highest LEP energies,\cite{OPALZZ} consistent with Ref.\cite{L3ZZold} It may therefore be prudent to await final results from L3 on this issue.} 
None of the experiments see any significant deviation from the SM in the analyzed channels. This is translated into the limits on $M_S({\rm Hewett})$, listed in Table~\ref{virtual}. They are of the order of 1~TeV for both signs of the interference term.

\begin{table*}[hbt]
\begin{center}
\caption{Lower limits at the 95\% CL on the effective Planck scale, $M_S({\rm Hewett})$, in TeV, from searches for direct graviton production at LEP. Limits from $\sqrt{s} > 200$~GeV data are shown in normal font; limits from 189~GeV data are in {\it italics\/}; limits from 184~GeV data are in {\bf bold} script.}
\label{direct}
\begin{tabular}{|l|ccccc|ccccc|}
\hline
{\footnotesize\rm Experiment}  & \multicolumn{5}{|c|}{$e^+e^- \to \gamma G_{\rm KK}$} &  
              \multicolumn{5}{c|}{$e^+e^- \to Z G_{\rm KK}$} \\
\cline{2-6}\cline{7-11}
		& $n$=2 & $n$=3 & $n$=4 & $n$=5 & $n$=6 & 
              $n$=2 & $n$=3 & $n$=4 & $n$=5 & $n$=6 \\
\hline
ALEPH & 1.10 & 0.86 & 0.70 & 0.60 & 0.52 & \bf 0.35 &  \bf 0.22 & \bf 0.17 & \bf 0.14 & \bf 0.12 \\
\hline
DELPHI& 1.25 & 0.97 & 0.79 & 0.68 & 0.59 & N/A  & N/A  & N/A  & N/A  & N/A  \\
\hline
L3	& \it 1.02 & \it 0.81 & \it 0.67 & \it 0.58 & \it 0.51 & \it 0.60 & \it 0.38 & \it 0.29 & \it 0.24 & \it 0.21 \\
\hline
OPAL	& \it 1.09 & \it 0.86 & \it 0.71 & \it 0.61 & \it 0.53 & N/A  & N/A  & N/A  & N/A  & N/A  \\
\hline
\end{tabular}
\end{center}
\end{table*}

\begin{table*}[hbt]
\begin{center}
\caption{Lower limits at the 95\% CL on the effective Planck scale, $M_S({\rm Hewett})$, in TeV, from searches for virtual graviton effect at LEP. Upper (lower) rows correspond to $\lambda=+1$ ($\lambda = -1$). The ALEPH Collaboration used a different formalism for their analysis,\protect\cite{GRW} so their limits were translated into Hewett's formalism.\protect\cite{Hewett} The L3 Collaboration used formalism\protect\cite{AD} for diboson production,\protect\cite{L3ZZold,L3ZZ} in which the sign of $\lambda$ is reversed, compared to Hewett.\protect\cite{Hewett} To correct for that, we reverse the sign of $\lambda$ when quoting the L3 limits in the $\gamma\gamma$, $WW$, and $ZZ$ channels. Combined L3 limits are nevertheless affected by the mixture of two signs of $\lambda$ in difermion and diboson channels. (See also footnote on the previous page for a discussion of the $ZZ$ results.) Limits from $sqrt{s} > 200$~GeV data are shown in normal font; limits from 189~GeV data are in {\it italics\/}; limits from 184~GeV data are in {\bf bold} script. Some of the older limits obtained within the formalism\protect\cite{GRW} before an important revision was made, are not directly comparable with the results at the highest LEP energies.}
\label{virtual}
\begin{tabular}{|l|ccccc|ccc|c|}
\hline
{\footnotesize\rm Experiment}  & $e^+e^-$ & $\mu^+\mu^-$ & $\tau^+\tau^-$ & $q\bar q \quad (b\bar b)$ & $f\bar f$ & $\gamma\gamma$ & $WW$ & $ZZ$ & {\footnotesize\rm Combined} \\
\hline
ALEPH & 0.81 & 0.67 & 0.62 & 0.57 (0.44) & 0.84 & 0.82 & N/A  & N/A  & \it 1.00 \\
	& 1.05 & 0.65 & 0.60 & 0.53 (0.44) & 1.05 & 0.81 & N/A  & N/A  & \it 0.75 \\
\hline
DELPHI& N/A  & 0.73 & 0.65 & N/A  (N/A)  & 0.76 & 0.71 & N/A  & N/A  & N/A \\
	& N/A  & 0.59 & 0.56 & N/A  (N/A)  & 0.60 & 0.69 & N/A  & N/A  & N/A \\
\hline
L3	& 0.99 & \it 0.69 & \it 0.54 & {\bf 0.49} (N/A)  & \it 1.00 & \it 0.80 & \it 0.68 & 1.2  & 1.3 \\
	& 0.91 & \it 0.56 & \it 0.58 & {\bf 0.49} (N/A)  & \it 0.84 & \it 0.79 & \it 0.79 & 1.2  & 1.2 \\
\hline
OPAL	& N/A  & \it 0.60 & \it 0.63 & N/A  (N/A)  & \it 0.68 & 0.82 & N/A  & 0.80 & 0.90 \\
	& N/A  & \it 0.63 & \it 0.50 & N/A  (N/A)  & \it 0.61 & 0.85 & N/A  & 0.59 & 0.83 \\
\hline
\end{tabular}
\end{center}
\end{table*}

Virtual graviton effects have also been sought at HERA in the $t$-channel of $e^\pm p \to e^\pm p$ scattering, similar to Bhabha scattering at LEP.\cite{GRW,Hewett} A search carried out by the H1 Collaboration\cite{H1} with 82~pb$^{-1}$ of $e^+p$ and 15~pb$^{-1}$ of $e^-p$ data, have set limits on $M_S$ between 0.5 and 0.8~TeV (see Table~\ref{H1}). Although these limits are somewhat inferior to those from LEP, the ultimate sensitivity of HERA at the end of the next run is expected to be similar to that at LEP.

Recently, the D\O\ Collaboration reported the first search for virtual graviton effects at a hadron collider,\cite{D0} based on the analysis of a two-dimensional distribution in the invariant mass and scattering angle of dielectron or diphoton systems, as suggested in Ref.\cite{KCGL} The results, corresponding to 127~pb$^{-1}$ of data collected at $\sqrt{s} = 1.8$~TeV, agree well with the SM predictions, and provided the limits on the effective Planck scale, shown in Table~\ref{D0} for all three formalisms.\cite{GRW,HLZ,Hewett} These limits are similar to and complementary to those from LEP, as different energy regimes are probed at the two colliders. A similar analysis in the dielectron channel is being pursued by the CDF Collaboration,\cite{Gerdes} but no results have yet been reported. As the current Tevatron sensitivity is limited by statistics, rather than machine energy, we expect combined Tevatron limits to yield an improvement over the currently excluded range of $M_S$.

\begin{table}[b]
\begin{center}
\caption{Lower limits at the 95\% CL on the effective Planck scale, $M_S{\rm Hewett}$, in TeV, from the H1 experiment.\protect\cite{H1} The limits have been translated into Hewett's formalism\protect\cite{Hewett} from the original formalism\protect\cite{GRW} used in the H1 analysis.}
\label{H1}
\begin{tabular}{|c|cc|c|}
\hline
H1 & $e^+p$ & $e^-p$ & Combined \\
\hline
$\lambda=+1$ & 0.45 & 0.61 & 0.56 \\
$\lambda=-1$ & 0.79 & 0.43 & 0.83 \\
\hline
\end{tabular}
\end{center}
\end{table}

Although no evidence for LED has been found so far, we are looking forward to the next generation of collider experiments to shed more light on the mystery of large extra dimensions. The sensitivity of the upgraded Tevatron experiments in the next run is expected to double (2~fb$^{-1}$) or even triple (15~fb$^{-1}$), which offers a unique opportunity to see LED effects in the next 5 years. The ultimate test of the theory of large extra dimensions will become possible at the LHC, where effective Planck scales as high as 10~TeV will be able to be probed.

\begin{table*}[htb]
\begin{center}
\caption{Lower limits at 95\% CL on the effective Planck scale, $M_S$, in TeV, from the D\O\ experiment.\protect\cite{D0}}
\label{D0}
\begin{tabular}{|c|@{}cccccc|@{}cc|}
\hline
GRW\cite{GRW} & \multicolumn{6}{@{}c|}{HLZ\cite{HLZ}} & \multicolumn{2}{@{}c|}{Hewett\cite{Hewett}} \\
\hline
& ~~$n$=2 & $n$=3 & $n$=4 & $n$=5 & $n$=6 & $n$=7~~ & ~~$\lambda=+1$ & $\lambda=-1$ \\
\cline{2-7} \cline{8-9}
1.21 & ~~1.37 & 1.44 & 1.21 & 1.10 & 1.02 & 0.97~~  & ~~1.08      & 1.01 \\
\hline
\end{tabular}
\end{center}
\end{table*}

\end{document}